\documentclass[preprint,showpacs,preprintnumbers,amsmath,amssymb,aps,prd]{revtex4-1}
\usepackage{graphics,epsfig,subfigure}
\usepackage{diagbox}
\usepackage[usenames]{color}
\usepackage[colorlinks,
            linkcolor=blue,
            anchorcolor=red,
            citecolor=red
            ]{hyperref}
\usepackage{color}
\usepackage{graphicx}
\usepackage{amsmath}
\usepackage{amsfonts}
\usepackage{amssymb}
\usepackage{txfonts}
\usepackage{indentfirst}
\usepackage{booktabs}
\begin{document}
\renewcommand{\baselinestretch}{1.3}
\newcommand\beq{\begin{equation}}
\newcommand\eeq{\end{equation}}
\newcommand\beqn{\begin{eqnarray}}
\newcommand\eeqn{\end{eqnarray}}
\newcommand\nn{\nonumber}
\newcommand\fc{\frac}
\newcommand\lt{\left}
\newcommand\rt{\right}
\newcommand\pt{\partial}

\title{Rotating multistate axion boson stars}   
\author{Yan-Bo Zeng, Shi-Xian Sun, Si-Yuan Cui, Yu-Peng Zhang and Yong-Qiang Wang\footnote{yqwang@lzu.edu.cn, corresponding author}
}


\affiliation{$^{1}$ Key Laboratory of Quantum Theory and Applications of MoE, Lanzhou University, Lanzhou 730000, China\\
$^{2}$Lanzhou Center for Theoretical Physics, Key Laboratory of Theoretical Physics of Gansu Province, School of Physical Science and Technology, Lanzhou University, Lanzhou 730000, China\\
$^{3}$Institute of Theoretical Physics $\&$ Research Center of Gravitation,
Lanzhou University, Lanzhou 730000, China}

\begin{abstract}
We consider excited configuration and multistate configuration of rotating axion boson stars~(RABSs).
RABSs are asymptotically flat, stationary, spinning, horizonless solutions of Einstein-Klein-Gordon theory in which the scalar potential depends on scalar field mass $\mu$ and axion decay constant $f_a$. 
The excited RABSs have two types of solutions, including $^2S$ state and $^2P$ state. 
The rotating multistate axion boson stars~(RMABSs) consist of coupled fundamental configuration and excited configuration, and also include two types of solutions: $^1S^2S$ state and $^1S^2P$ state. 
Some differences between RABSs models and rotating mini-boson stars models are discussed. 
We show the solution space of these models for different values of decay constant $f_a$. 
We found fundamental RABSs have a higher maximum mass than the excited state in the low decay constant region. 
Moreover, the multistate configuration allows a higher mass than both the fundamental configuration and the excited configuration at the same frequency. 
We also found the RMABSs have the second branch in which the ergo-region emerges. 
This means the second branch may be superradiant unstable. 
\end{abstract}

\maketitle

\section{Introduction}\label{Sec1}
The concept of boson stars is from the study of geons, proposed by Wheeler~\cite{Wheeler:1955zz,Power:1957zz}. 
The geons are considered as stable, particle-like, and horizonless solutions in the context of gravity coupled to a classical electromagnetic field. 
However, the geons consisting of massless vector field are not found. 
Later, Kaup~{\em et al.} obtained Klein-Gordon geons~({\em i.e.} boson stars) by replacing massless vector field with massive complex scalar field~\cite{Kaup:1968zz}. 
Ruffini also independently studied boson stars by considering quantized real scalar field~\cite{PhysRev.187.1767}. 
The original boson stars are the spherical, free scalar, and fundamental configurations. 
It is generalized to the cases of rotation~\cite{Schunck:1996,Schunck:1996he,1997PhRvD..56..762Y}, the excited BSs~\cite{Bernal:2009zy,Collodel:2017biu,Wang:2018xhw}, static multipolar BSs~\cite{Herdeiro:2020kvf,Herdeiro:2021mol}, the construction of vector boson stars~(Proca stars) - see~{\em e.g.}~\cite{Brito:2015pxa,Herdeiro:2019mbz,Minamitsuji:2018kof} and different kinds of multi-field multi-state configuration~\cite{Bernal:2009zy,Alcubierre:2018ahf,Li:2019mlk,Li:2020ffy,Sanchis-Gual:2021edp,Henriques:1989ez,Zeng:2021oez,Dzhunushaliev:2021vwn}. 
There are other generalizations as well, see the reviews~\cite{Schunck:2003kk,Liebling:2012fv}. 

Repulsive forces arising in Heisenberg's uncertainty principle and gravity form a balance, 
which causes that the smaller the mass of bosons, the larger the maximum mass of boson stars. 
Thus, boson stars consisting of ultralight boson particles are expected to be astrophysical, macroscopic objects. 
These objects may be considered as black hole mimickers~\cite{Feinblum:1968nwc,Kaup:1968zz,PhysRev.187.1767} and dark matter lumps~\cite{Sahni:1999qe,Hu:2000ke,Matos:2000ng}. 
Theoretically, axion-like particles~(ALPs) are one of the most motivated hypothetical ultralight bosons. 
The studies~\cite{Weinberg:1977ma,Wilczek:1977pj,Peccei:1977hh} of Quantum Chromo Dynamics~(QCD) axion was motivated by solving the strong CP problem~\cite{Peccei:1977hh,Jackiw:1976pf} in 1977. 
Moreover, ALPs with similar properties, which are extremely light and weakly-interacting, are naturally considered in string theory~\cite{Jaeckel:2010ni,Goodsell:2009xc,Arvanitaki:2010sy}. 
Usually, axion is also seen as one of the potential dark matter candidate~\cite{Peccei:1977hh,Weinberg:1977ma,Wilczek:1977pj}. 
Currently, the nodeless fundamental configuration of axion boson stars, including spherical cases and rotating cases, are studied in Refs.~\cite{Guerra:2019srj,Delgado:2020udb}. 
These studies show that the properties of rotating axion boson stars are strongly related to the self-interaction of axion potential, which is characterized by the inverse of axion decay constant $f_a$. 
If $f_a$ is large~({\em i.e.} $f_a=1$), the axion stars models matche the mini-boson stars models. 
For sufficiently small decay constants, the axion stars have different features, including ADM mass, compactness, and stability. 

The goal of this paper is to generalize the excited configuration and multistate configuration of rotating mini-boson stars to the rotating axion boson stars. 
The rotating multistate axion boson stars are composed of two axion fields, which are in the ground state and the first excited state, respectively. 
Then, we study the influence of decay constant $f_a$ on these configurations under the synchronized frequency condition and the nonsynchronized frequency condition, 
and plot the distribution of the axion field and the parameter space of solutions. 

This paper is organized as follows: In Sec.~\ref{sec2}, we introduce the model of rotating multistate axion boson stars, including the motion equations, the metric ansatz and axion field ansatz. 
In Sec.~\ref{sec3}, we show boundary conditions of ansatz functions and the formula of interest quantities. 
In Sec.~\ref{sec4}, we introduce numerical approach and give two subsections: \ref{1s2s} and \ref{1s2p}. 
Two subsections show the numerical results of $^1S^2S$ state and $^1S^2P$ state. 
In Sec.~\ref{sec5}, conclusions and perspectives are exhibited.

\section{The model setup}\label{sec2}
Here, we mainly introduce the model of the RMABSs. 
In order to describe the system of axion field minimally coupled to Einstein gravity, we start with the EinsteinKlein-Gordon (EKG) action, 
\begin{equation}
  \label{action}
  S=\int\sqrt{-g}d^4x\left(\frac{R}{16\pi G}+\mathcal{L}_{m}\right) \ ,
\end{equation}
where $R$ corresponds to the Ricci curvature and $G$ is Newton’s constant. $\mathcal{L}_{m}$ denotes the Lagrangian of the matter field, for double scalar fields~($i=0,1$), which reads, 
\begin{equation}
  \label{lag}
  \mathcal{L}_{m}= \sum -\nabla_{\mu}\psi_i^*\nabla^{\mu}\psi_i - U_i(|\psi_i|^2), \hspace{5pt} i=0,1 \ .
\end{equation}
From the action~(\ref{action}), we can obtain the EKG equation, 
\begin{equation}
\label{eq:einstein}
E_{\mu\nu}=R_{\mu\nu}-\frac{1}{2} g_{\mu\nu} R-8 \pi T_{\mu\nu}=0 \ ,
\end{equation}
\begin{equation}
\label{eq:scalar}
\square \psi_{i}-\frac{\partial U_{i}}{\partial\left|\psi_{i}\right|^{2}} \psi_{i}=0, \hspace{5pt} i=0,1 \ .
\end{equation}
Here Eq.~(\ref{eq:einstein}) is Einstein field equations, in which $T_{\mu\nu}$ is stress-energy tensor of two axion fields. 
Eq.~(\ref{eq:scalar}) governs the axion field dynamics. 

The metric of rotating boson star can be expressed as
\begin{eqnarray}
  d s^{2}=-e^{2 F_{0}} d t^{2}+e^{2 F_{1}}\left(d r^{2}+r^{2} d \theta^{2}\right)+e^{2 F_{2}} r^{2} \sin ^{2} \theta(d \varphi-W d t)^{2} \ .
\end{eqnarray}
Here $F_0(r,\theta)$, $F_1(r,\theta)$, $F_2(r,\theta)$ and $W(r,\theta)$ are specified ansatz functions. 
Furthermore, we assume stationary axion fields in the form, 
\begin{eqnarray} 
  \psi_i&=\phi_{i}(r,\theta)e^{i(m_i\varphi-\omega_i t)}, \hspace{5pt} m_i=\pm1,\pm2,\cdots \hspace{5pt} i=0,1 \ .
\end{eqnarray}
Here, $\psi_0$ is the axion field in the ground state,
and the other axion field $\psi_1$ is in the first excited state, respectively. 
$m_i$ and $\omega_i$ represent the azimuthal harmonic index and axion field frequency. 
As we have done in our previous work, both axion fields have the same azimuthal harmonic index, 
and we only set $m_0=m_1=1$. 
In addition, if the two axion fields have the same field frequency~({\em i.e.} $\omega_0=\omega_1$), this case is in the synchronized frequency condition. 
If the two axion fields have different field frequencies, this case is in the nonsynchronized frequency condition. 

In the following, we specify the form of the potential of two axion fields, 
\begin{eqnarray}
  U_i=\frac{2\mu_i^2f_a^2}{B}\left[1-\sqrt{1-4B \sin^2	\left(\frac{\sqrt{|\psi_i|^2}}{2f_a}\right)}\right], \hspace{5pt} i=0,1 \ .
\end{eqnarray}
Here $B$ depends on the masses of the up and down quarks, $B \approx 0.22$. 
From the above equation, it can be seen that two axion fields have the same axion decay constant $f_a$. 
Notice that we did not find the solutions of multistate axion stars in which two fields have the same parameters. 
Thus, we consider the two fields have different axion field mass. 
Moreover, by expanding the axion potential around $\psi_i = 0$~($i=1,2$), we obtain
\begin{eqnarray}
U_i(|\psi_i|^2)=\mu_i^2|\psi_i|^2-\left(\frac{3B-1}{12}\right)\frac{\mu_i^2}{f_a^2}|\psi_i|^4+\cdots \ .  \label{expansion}
\end{eqnarray}
According to the above equation, it is convenient to consider $f_a$ as a measure of self-interaction. 
When axion decay constant $f_a$ is much larger than these two axion fields $\psi_0$ and $\psi_1$, 
higher-order terms can be ignored in Eq.~(\ref{expansion}). 
So the behavior of rotating multistate axion boson stars will reduce to rotating multistate boson stars without self-interaction.

\section{Boundary conditions}\label{sec3}
To solve the EKG equation in asymptotically flat spacetime, 
we give appropriate boundary conditions for ansatz functions as follows
\begin{eqnarray}
F_0=F_1=F_2=W=\phi_{0}=\phi_{1}=0 \hspace{5pt}
\end{eqnarray}
at infinity~($r \rightarrow \infty$). Due to the axial symmetry of solutions, we require 
\begin{equation}\label{abc}
\partial_\theta F_0=\partial_\theta F_1=\partial_\theta F_2 =\partial_\theta W=\phi_{0}=\phi_{1}=0
\end{equation}
on the positive $z$ axis ($\theta=0$). 
It is convenient to consider the range $0 \leq \theta \leq \pi/2$. 
On the one hand, we can reduce the computational effort by specifying the reflection symmetry across the equatorial plane. 
On the other hand, 
the symmetry of ansatz functions about the equatorial plane can be used to distinguish between two types of solutions. 
For even parity solutions, we require the following boundary conditions 
\begin{eqnarray}
  \partial_\theta F_0=\partial_\theta F_1=\partial_\theta F_2 = \partial_\theta W=\partial_\theta \phi_{0} =\partial_\theta \phi_{1} = 0
\end{eqnarray}
in the equatorial plane~($\theta=\pi/2$). 
For odd parity solutions, we require
\begin{eqnarray}
  \partial_\theta F_0=\partial_\theta F_1=\partial_\theta F_2 = \partial_\theta W=\phi_{0} = \phi_{1} = 0 
\end{eqnarray}
in the equatorial plane. 
At the origin ($x=0$), we impose 
\begin{align}
\phi_{0}&=\phi_{1} = 0 \ , \nonumber \\
\partial_r F_0=\partial_r F_1&=\partial_r F_2=\partial_r W = 0 \ .
\end{align}

The ADM mass $M$ and angular momentum $J$ are the key quantities we are interested in. 
These quantities are encoded in the asymptotic expansion of metric components, see~{\em e.g.}~\cite{Grandclement:2014msa}, 
\begin{eqnarray}
g_{tt}= -1+\frac{2GM}{r}+\cdots \ , \nonumber\\
g_{\varphi t}= -\frac{2GJ}{r}\sin^2\theta+ \cdots \ .
\end{eqnarray}
In the Kerr spacetime, the presence of the ergo-region is a key feature for black holes and ultracompact objects. 
When we set the signature convention as $(-, +, +, +)$ for the metric, ergo-region are defined as a spacetime region in which the $g_{tt}$ is positive. 
The ergo-surface has the following form, 
\begin{eqnarray}
g_{t t}=-e^{2 F_{0}}+W^{2} e^{2 F_{2}} \sin ^{2} \theta=0 \ .
\end{eqnarray}

\section{Numerical results}\label{sec4}
In this work, all the numbers are dimensionless as follows
\begin{eqnarray}
r \rightarrow r\mu_0 \hspace{5pt}, \hspace{5pt} \phi \rightarrow \phi M_{Pl} \hspace{5pt}, \hspace{5pt} \omega_i \rightarrow \omega_i/\mu_0 \hspace{5pt}, \hspace{5pt} \mu_1 \rightarrow \mu_1/\mu_0 \ .
\end{eqnarray}
Here $M_{Pl}=G^{-1/2}$ is the Plank mass. 
We set $G = c = \mu_0 = 1$ and $\mu_1=0.93$ for easy comparison with those of previous works. 
We transform the radial coordinates by the following equation, 
\begin{eqnarray}
\label{transform}
x=\frac{r}{1+r} \ .
\end{eqnarray}
The use of Eq.~(\ref{transform}) map the infinite region $[0,\infty)$ to the finite region $[0,1)$. 
Next, what we need is to discretize the EKG equations on a grid. 
This allows the partial differential equations to be approximated by algebraic equations. 
The grid with $250 \times 160$ points covers the integration region $0 \leq x \leq 1$ and $0 \leq \theta \leq \pi/2$. 
No significant effect of grid density on physical quantities. 
In our work, relative errors are less than $10^{-6}$. 

Due to the similarity of Eq.~(\ref{eq:scalar}) and Schrödinger equation, boson stars are also known as macroscopic atoms. 
Analogous to the hydrogen orbital, 
the ground state~({\em i.e.} nodeless scalar field, $n=0$) is called as $^1S$ state. 
For the first excited state~($n=1$), boson stars have two types of $n_r=1$ and $n_\theta=1$ solutions. 
Here $n_r$ is the number of nodes along the radial direction, and $n_\theta$ is the number of nodes along the angular direction. 
The kind of solutions with $n_r=1$ are $^2S$ state. 
Instead, the kind of solutions with $n_\theta=1$ are $^2P$ state. 
The simplest non-trivial configuration with two complex scalar fields was discussed in Ref.~\cite{Li:2019mlk}. 
The multistate configuration of two scalar fields, in which one is in the $^1S$ state~($\mu_0=1$) and the other is in the $^2S$ state~($\mu_1=0.93$), is named the $^1S^2S$ state. 
Besides, the multistate configuration with the $^1S$ state~($\mu_0=1$) and the $^2P$ state~($\mu_1=0.93$) is named the $^1S^2P$ state. 
Here, 
we mainly analyze the ADM mass $M$ and the angular momentum $J$ under the synchronized frequency condition and the nonsynchronized frequency condition. 
In the nonsynchronized frequency condition, we fix the frequency of the axion field in the $^1S$ state, $\omega_0=0.8$. 
To study the properties of the ergo-surface, we also calculated $g_{tt}$ for the ground state, the excited state, and the multistate in the case of synchronized frequency. 
Previous studies have shown that fundamental axion stars have more branches as axion decay constant $f_a$ decreases~\cite{Guerra:2019srj,Delgado:2020udb}. 
We expect that the excited configuration and multistate configuration of rotating axion stars have similar behavior.  
In order to study the effect of self-interactions on the model, 
we choose the following values for the decay constant $f_a$, 
\begin{eqnarray}
f_{a}=\{1, 0.025, 0.015, 0.009\} \ ,
\end{eqnarray}
and study corresponding solutions. 
The case with $f_a=1$ corresponds to rotating mini-boson stars~(mini-BSs) without self-interaction. 
As $f_a$ decreases, self-interaction is stronger. 
In this model, the solution with $f_a=0$ is so complex that we can not obtain it. 
\clearpage

\subsection{\texorpdfstring{$^2S$ state and $^1S^2S$ state}{1S2S state}} \label{1s2s}

\begin{figure}[h!]
\begin{center}
\subfigure{\includegraphics[height=0.25\textheight]{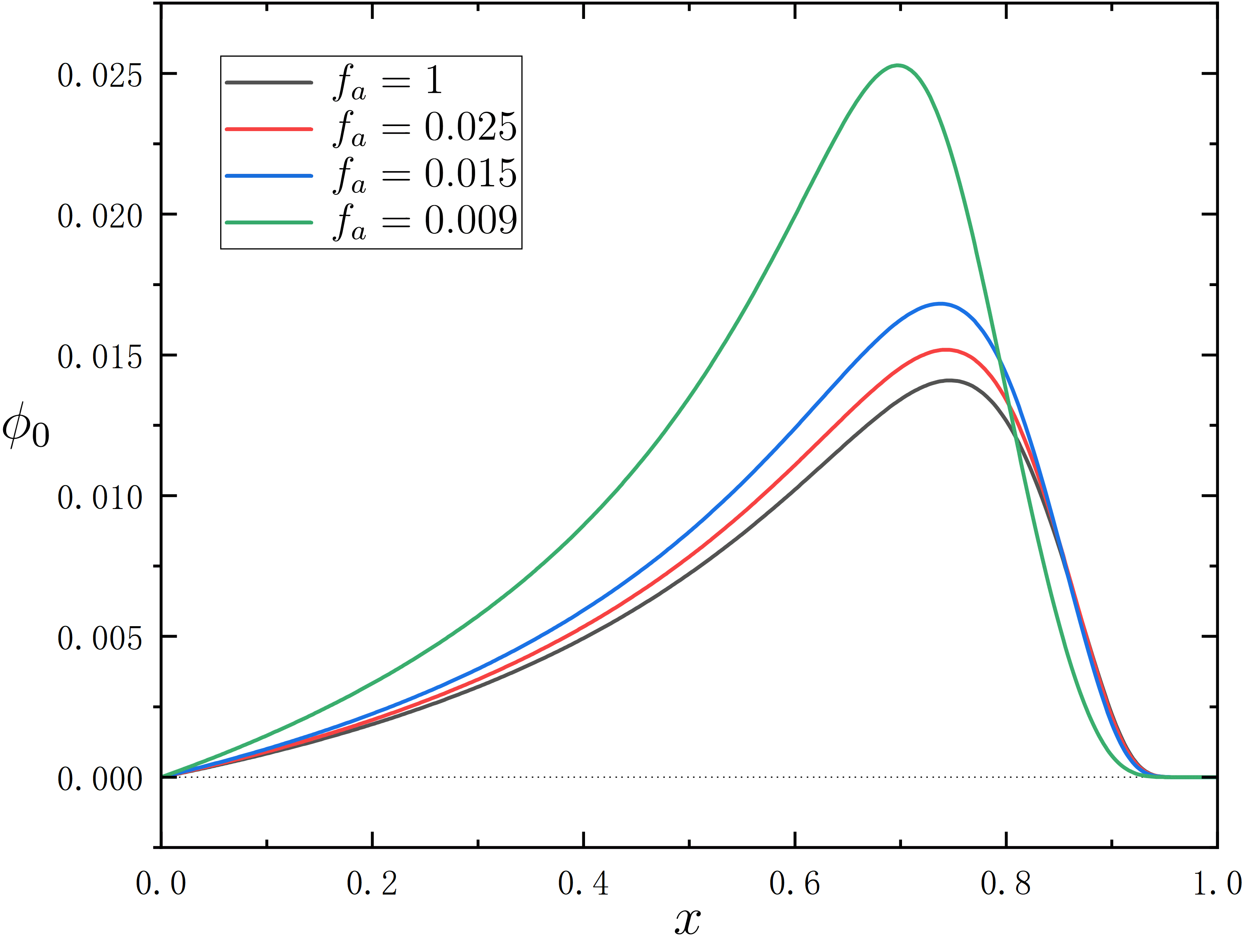} }
\subfigure{\includegraphics[height=0.25\textheight]{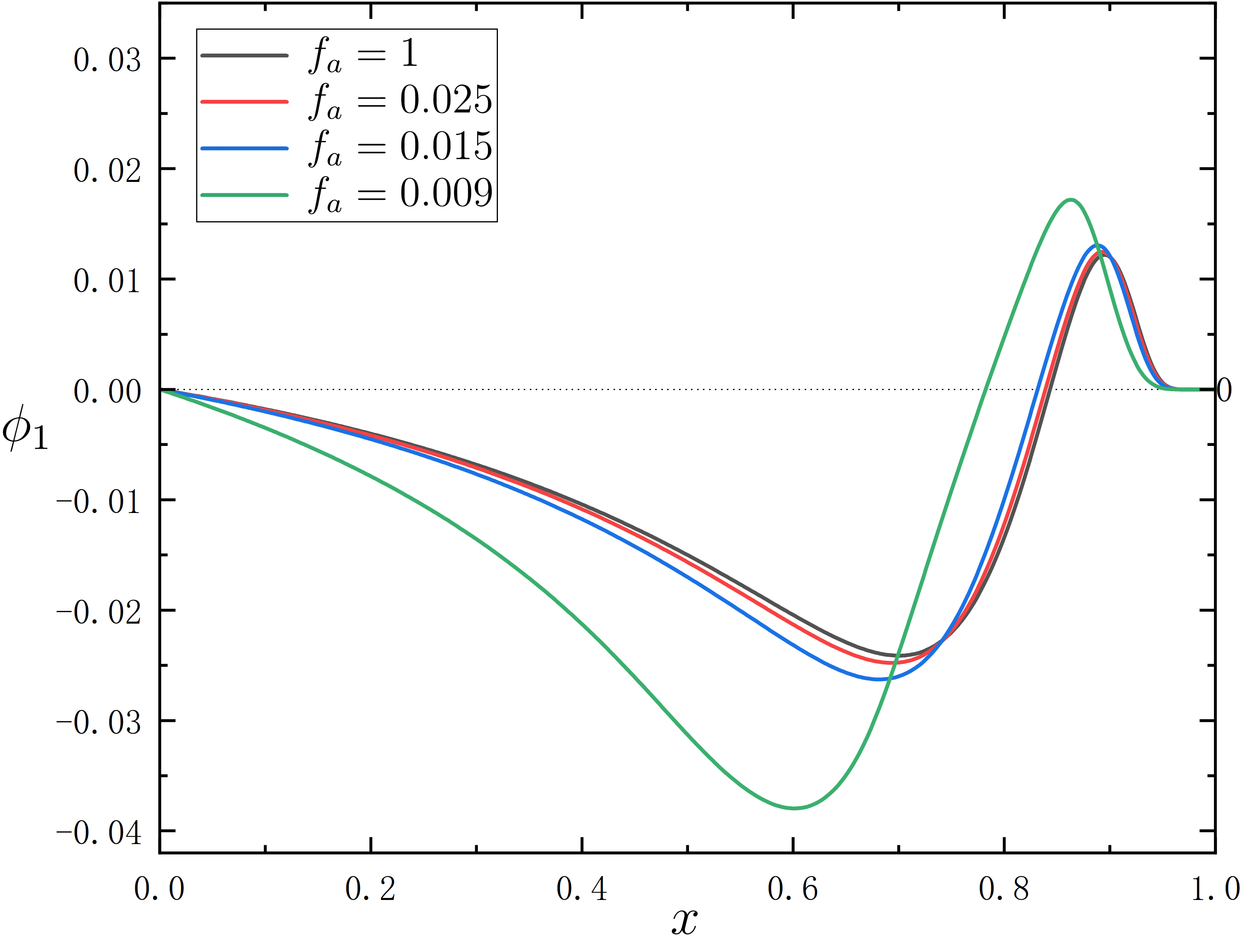} } 
\end{center}
\caption{The radial distribution of two axion fields in the RMABSs with the $^1S^2S$ state at $\theta=\pi/2$. \textit{Left}: The distribution of the axion field $\phi_0$ in the ground state $^1S$ state as a function of $x$ and $\theta$ for the frequency $\omega_0=0.8$. 
\textit{Right}: The distribution of the other axion field $\phi_1$ in the excited state $^2S$ state as a function of $x$ and $\theta$ for the frequency $\omega_1=0.8$.}
\label{fig:1s2s-distribution}
\end{figure}

As an example of rotating axion stars in $^1S^2S$ state, Fig.~\ref{fig:1s2s-distribution} shows the radial distributions of two axion fields $\phi_0$ and $\phi_1$ for four different axion decay constant $f_a$. 
Along the radial $r$ direction, the axion field in the ground state $^1S$ has no node. 
The axion field in the first excited state $^2S$ has one node. 
As $f_a$ decreases, $\phi_0$ and $\phi_1$ have higher maximum, 
and the peak value corresponds to a smaller radius. 
This means that RMABSs become more compact. 

\begin{figure}[h!]
  \centering
      \includegraphics[width=0.9\textwidth]{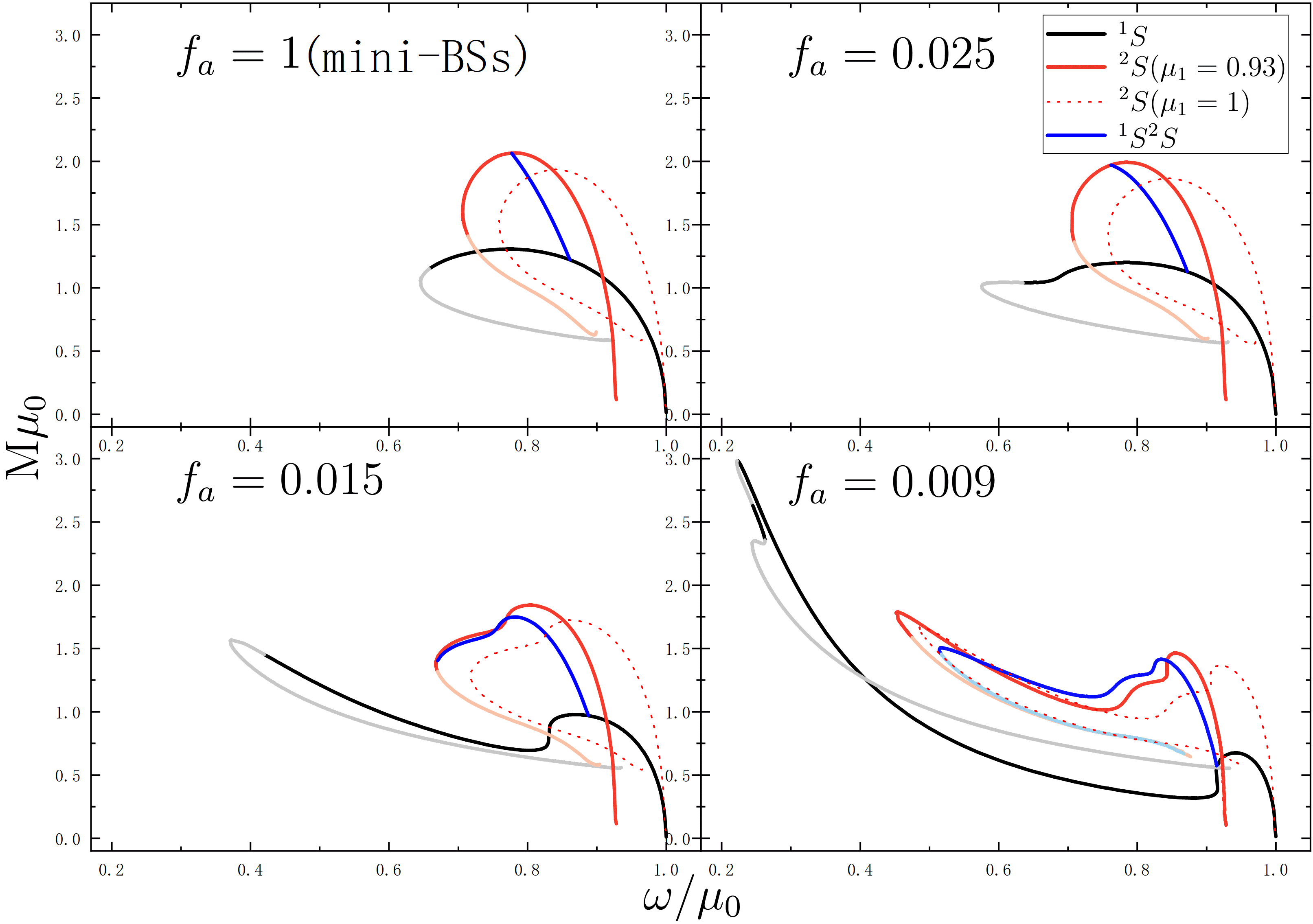}
      \caption{The ADM mass of the rotating axion stars as a function of the axion field frequency. The black line indicates $^1S$ state, the red solid line indicates the $^2S$ state with $\mu_1=0.93$, the red dotted line indicates the $^2S$ state with $\mu_1=1$, and the blue line indicates $^1S^2S$ state in the case of synchronized frequency, respectively. The light-colored line segment represents the emergence of ergo-region. }
      \label{M1s2s-sycronize-M}
\end{figure}

For the $^1S$ state, $^2S$ state and $^1S^2S$ state under the condition of synchronized frequency, the relation of $M-\omega$ is given in Fig.~\ref{M1s2s-sycronize-M}. 
We can see the ADM mass $M$ versus axion field frequency $\omega$ distribution for $^1S^2S$ state with four different values of $f_a$. 
The ground state $^1S$, the excited state $^2S$ with $\mu_1=0.93$, and the coexisting state $^1S^2S$ with $\mu_1=0.93$ represent the black, solid red, and blue lines, respectively. 
To compare the behavior of the ground state $^1S$ and the excited state $^2S$ in the same parameters, we also draw the dotted red line which indicates the excited state $^2S$ with $\mu_1=1$. 
The light-colored line segment represents the emergence of the ergo-surface. 
The fundamental rotating axion stars were first studied in Ref.~\cite{Delgado:2020udb}. 
The ground state $^1S$ with $f_a=1$ matches the fundamental solutions of rotating mini-boson stars, which form a spiraling curve. 
The mass $M$ increases from the Newtonian limit of $\omega_0\rightarrow 1$ until the global maximum of order $M_{Pl}^2/\mu_0$, 
and then the curve seems to spiral to the center~(We ignore the third branch). 
It is seen from the figure that, as $f_a$ decreases, the behavior of the black line deviates from the spiral line. 
The original global maximum of mass becomes the local maximum. 
The existence domain of frequency is extended. 
The minimum frequency $\omega_{0,min}$ gets smaller, near which corresponds to a global maximum mass if $f_a$ is particularly small. 
We find that the above features also appear in the case of the excited state $^2S$. 
Let's compare the excited state $^2S$~(dotted red line) with the ground state $^1S$~(black line). 
For the model of rotating mini-boson stars~($f_a=1$), the $^2S$ state has a higher mass than the $^1S$ state, regardless of the comparison of the same frequency or maximum mass. 
On the contrary, the fundamental RABSs have a higher maximum mass than RABSs in the $^2S$ state when $f_a$ is sufficiently small. 
The solid red line is the $^2S$ state solutions with $\mu_1=0.93$, which starts from the point of $\omega_1=0.93$~(Bound state condition requires $\omega_1 \leq \mu_1$). 
The $^2S$ state solutions with $\mu_1=0.93$ has a higher maximum mass and a lower minimum frequency than the case of $\mu_1=1$. 
Next, in the case of synchronized frequency condition, we analyze the RMABSs in the $^1S^2S$ state. 
Analogous to the multistate configuration of rotating mini-boson stars, the blue line also connects the black line to the solid red line. 
When $f_a$ is large, the RMABSs reduce to rotating mini-boson stars, and the mass of $^1S^2S$ state is higher than the ground state $^1S$ and smaller than the first excited state $^2S$. 
The corresponding curve has a monotonic behavior. 
As $f_a$ decreases, the blue line presents some characteristics of the solid red line, including the expansion of the existence domain and the existence of the second branch. 
For the cases of $f_a=0.009$, the $^1S^2S$ state has a higher mass than both the $^1S$ state and the $^2S$ state at the same frequency in some regions. 
We also studied the ergo-region of these configurations, which will be discussed later in subsection~\ref{1s2p}. 

\begin{figure}[h!]
  \centering
      \includegraphics[width=0.9\textwidth]{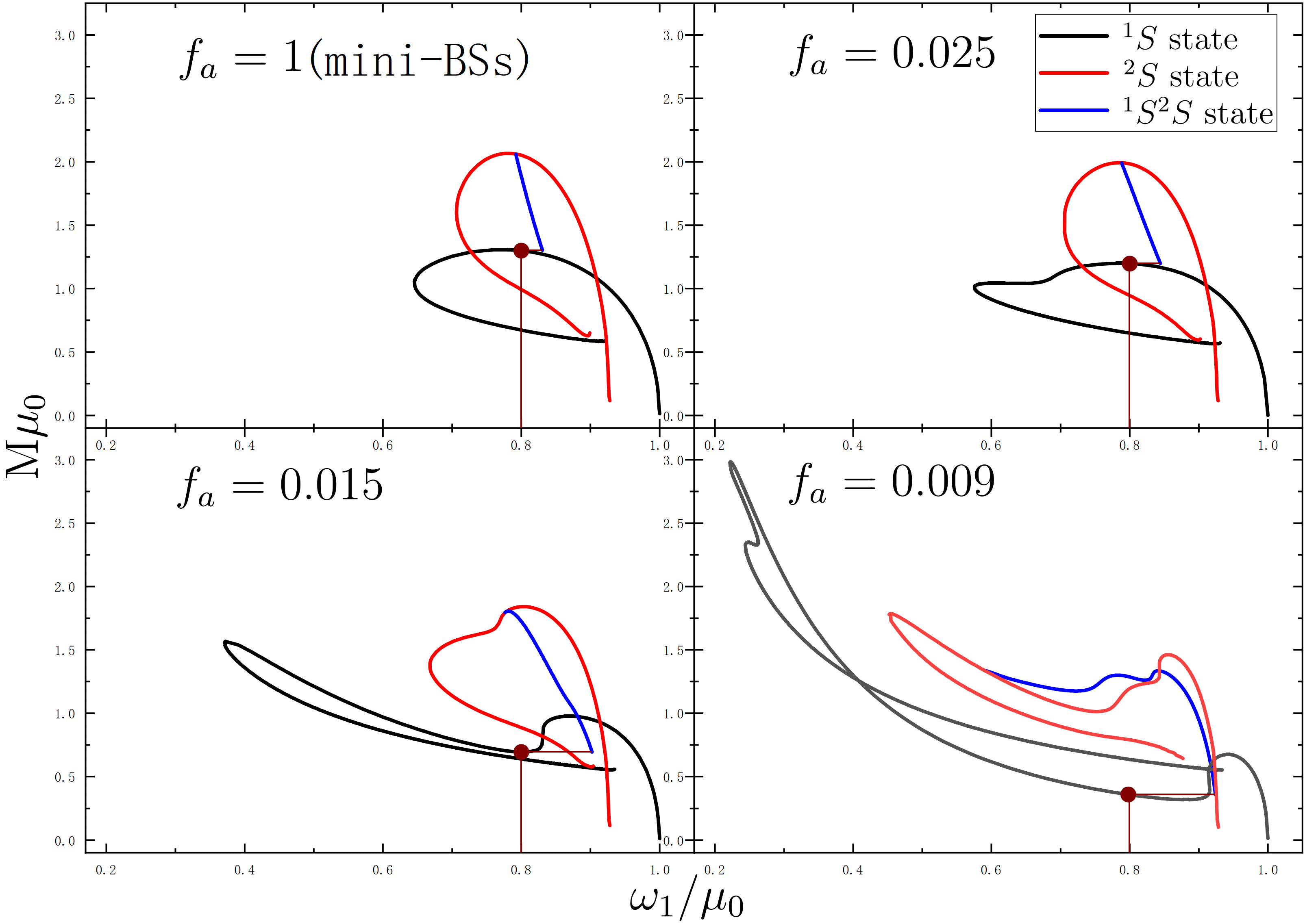}
      \caption{The ADM mass of the rotating axion stars as a function of the axion field frequency. The black line indicates $^1S$ state, the red solid line indicates the $^2S$ state with $\mu_1=0.93$, the red dotted line indicates the $^2S$ state with $\mu_1=1$, and the blue line indicates $^1S^2S$ state in the case of nonsynchronized frequency, respectively.}
      \label{M1s2s-nonsycronize-M}
\end{figure}

In the nonsynchronized frequency condition, we let $\omega_0=0.8$. 
We present the mass $M$ versus the nonsynchronized frequency $\omega_1$ in Fig.~\ref{M1s2s-nonsycronize-M}. 
As $\omega_1$ tends to the minimum, the RMABSs reduce to the $^2S$ state. 
As $\omega_1$ tends to the maximum, the blue line ends at a point, which corresponds to the solution of the $^1S$ state with $\omega_0=0.8$. 
Compared to the case of synchronized frequency, 
the shape of the blue line is still complicated for low $f_a$. 
However, there is no second branch. 
The minimum of $\omega_1$ is not as small as the case of synchronized frequency, which may cause the second branch not to exist. 

\begin{figure}[h!]
\begin{center}
\subfigure{\includegraphics[height=0.25\textheight]{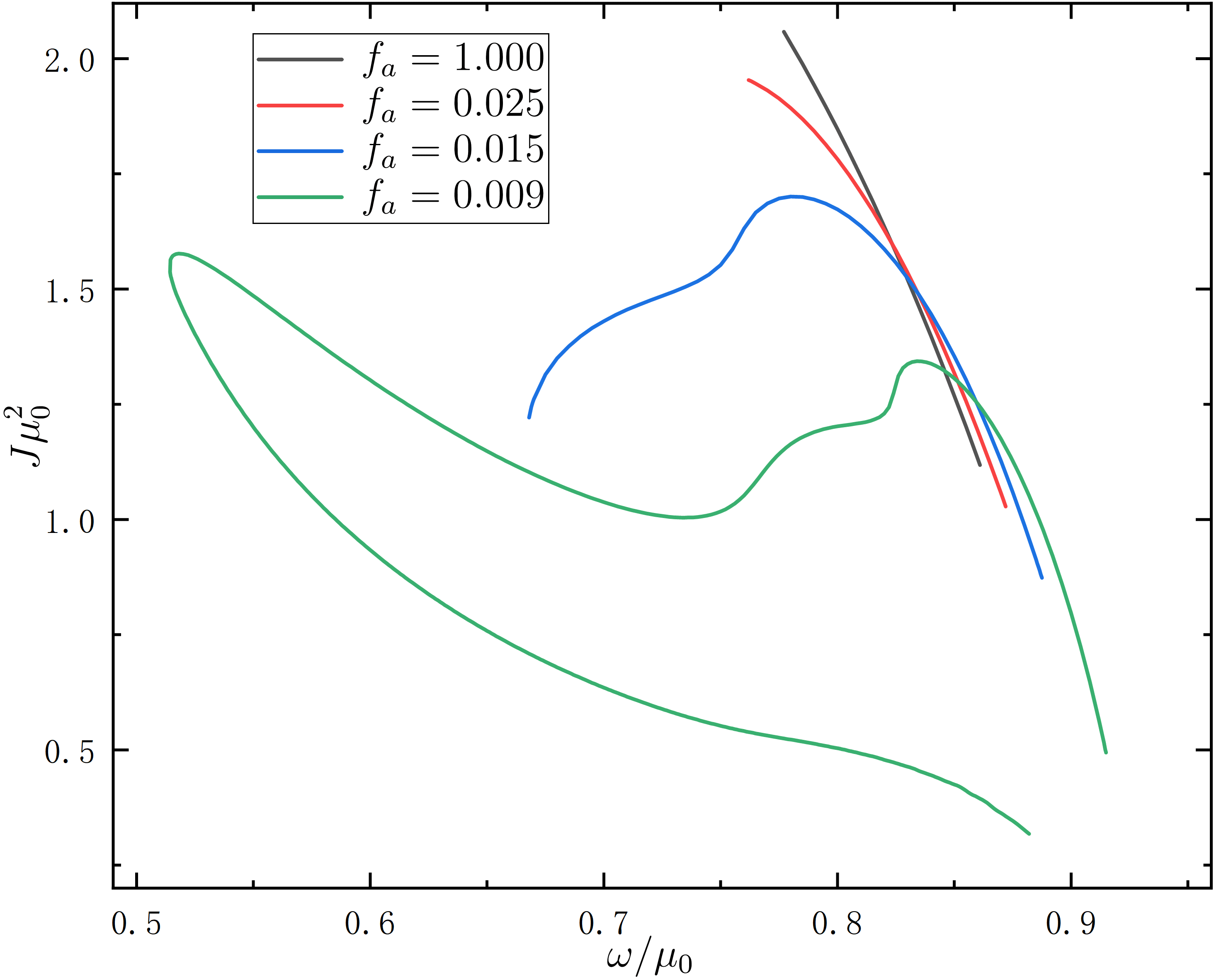} }
\subfigure{\includegraphics[height=0.25\textheight]{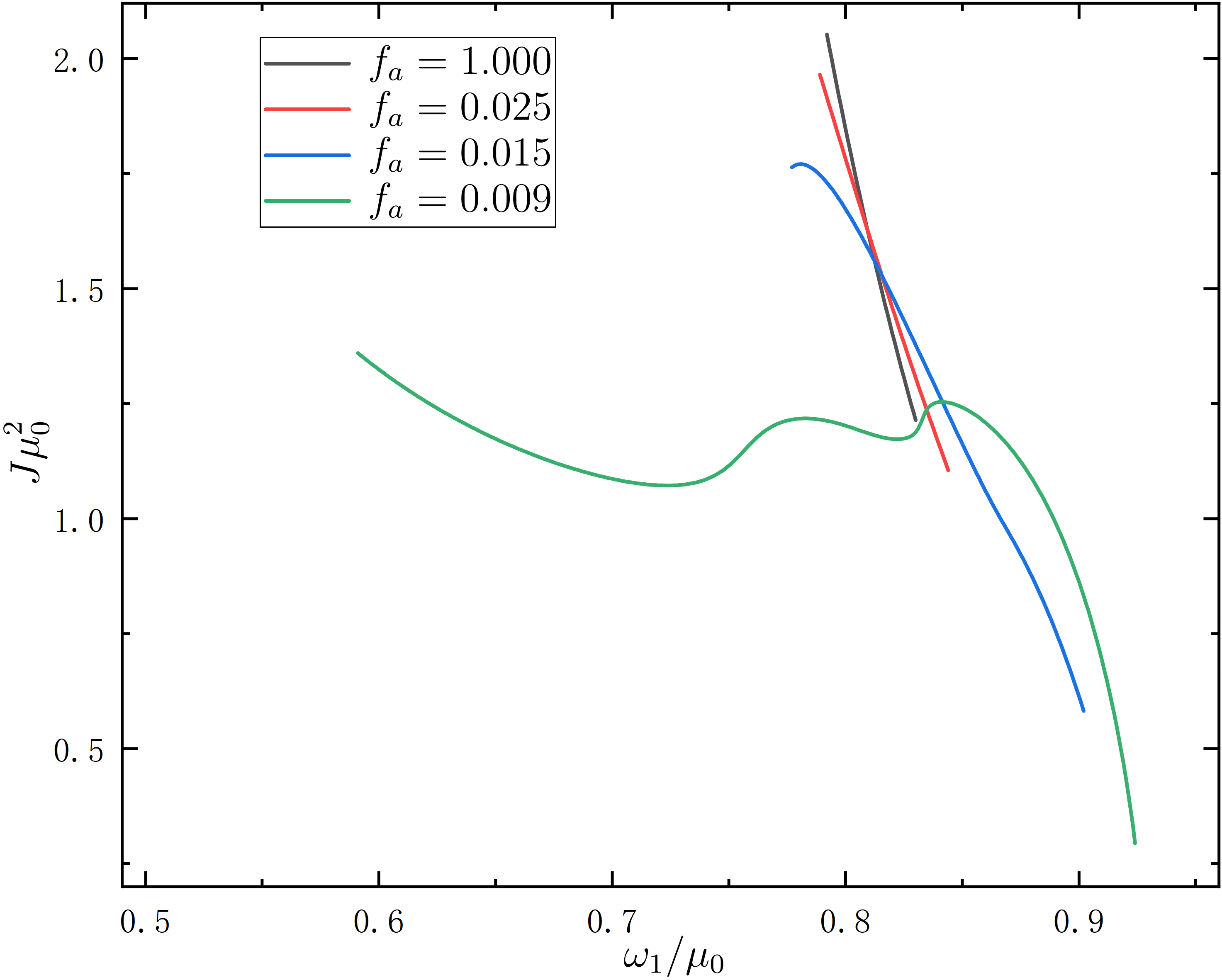} } 
\end{center}
\caption{The angular momentum of the RMABSs versus the axion field frequency for the coexisting state $^1S^2S$ with $f_a=\{1,0.025,0.015,0.009\}$. \textit{Left}: The RMABSs under the conditions of synchronized frequency. \textit{Right}: The RMABSs under the conditions of nonsynchronized frequency~($\omega_0=0.8$).}
\label{fig:1s2s-J}
\end{figure}
In Fig.~\ref{fig:1s2s-J}, we compared the behavior of angular momentum $J$ of the RMABSs with the coexisting state $^1S^2S$ under the conditions of synchronized frequency and nonsynchronized frequency. 
On the left panel, the angular momentum of the $^1S^2S$ state with $f_a=1$ and $f_a=0.025$ versus synchronized frequency $\omega$ is monotonic. 
When $f_a$ becomes smaller, the maximum angular momentum becomes lower, and the angular momentum as a function of axion field frequency is no longer monotonic. 
Then, the second branch emerges like $M$-$\omega$ diagram. 
On the right panel, something is different. 
The case of $f_a=0.015$ is monotonic. 
Moreover, the case of $f_a=0.009$ does not show the second branch because of the reduced frequency range.

\subsection{\texorpdfstring{$^2P$ state and $^1S^2P$ state}{1S2P state}} \label{1s2p}

\begin{figure}[h!]
\begin{center}
\subfigure{\includegraphics[height=0.25\textheight]{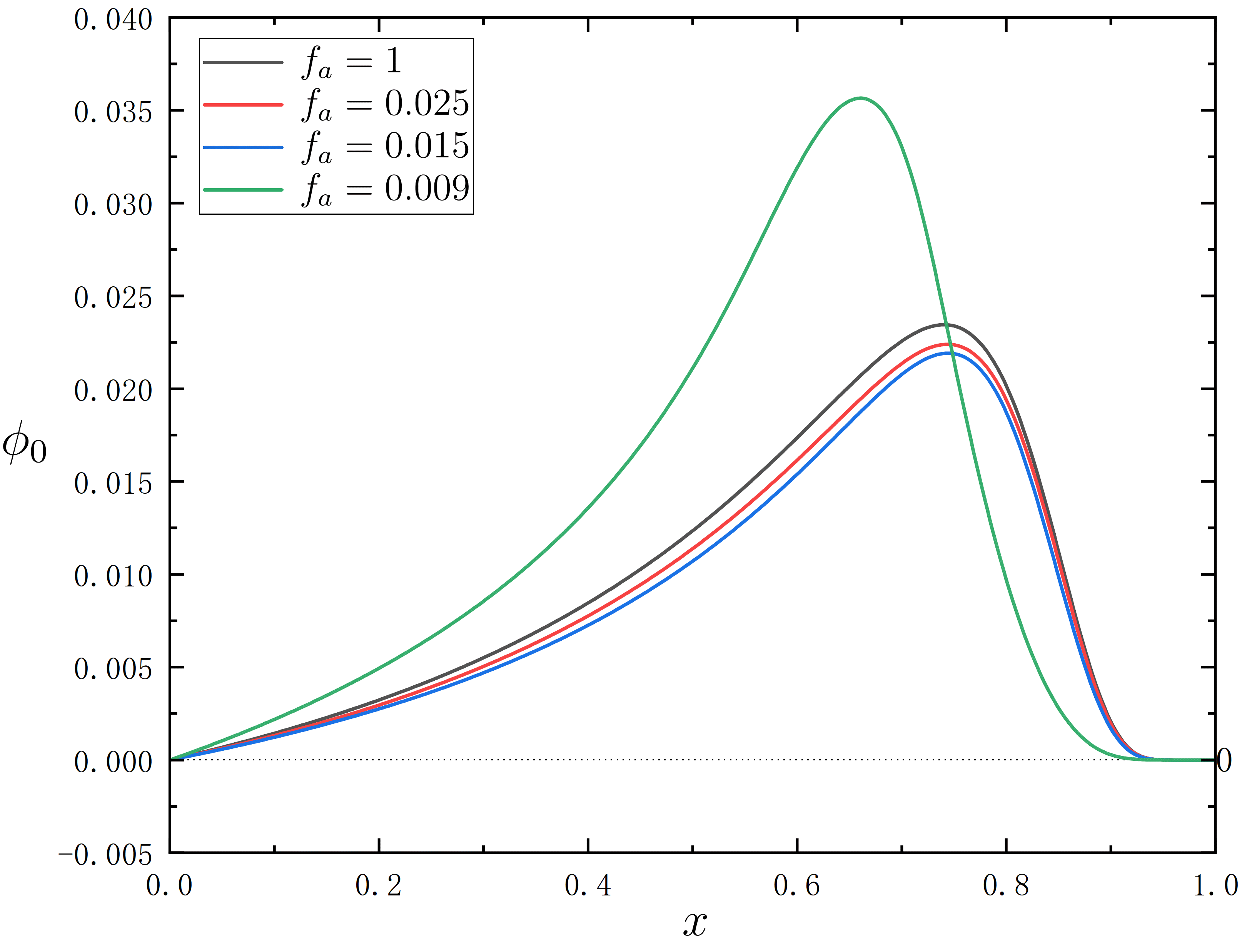} }
\subfigure{\includegraphics[height=0.25\textheight]{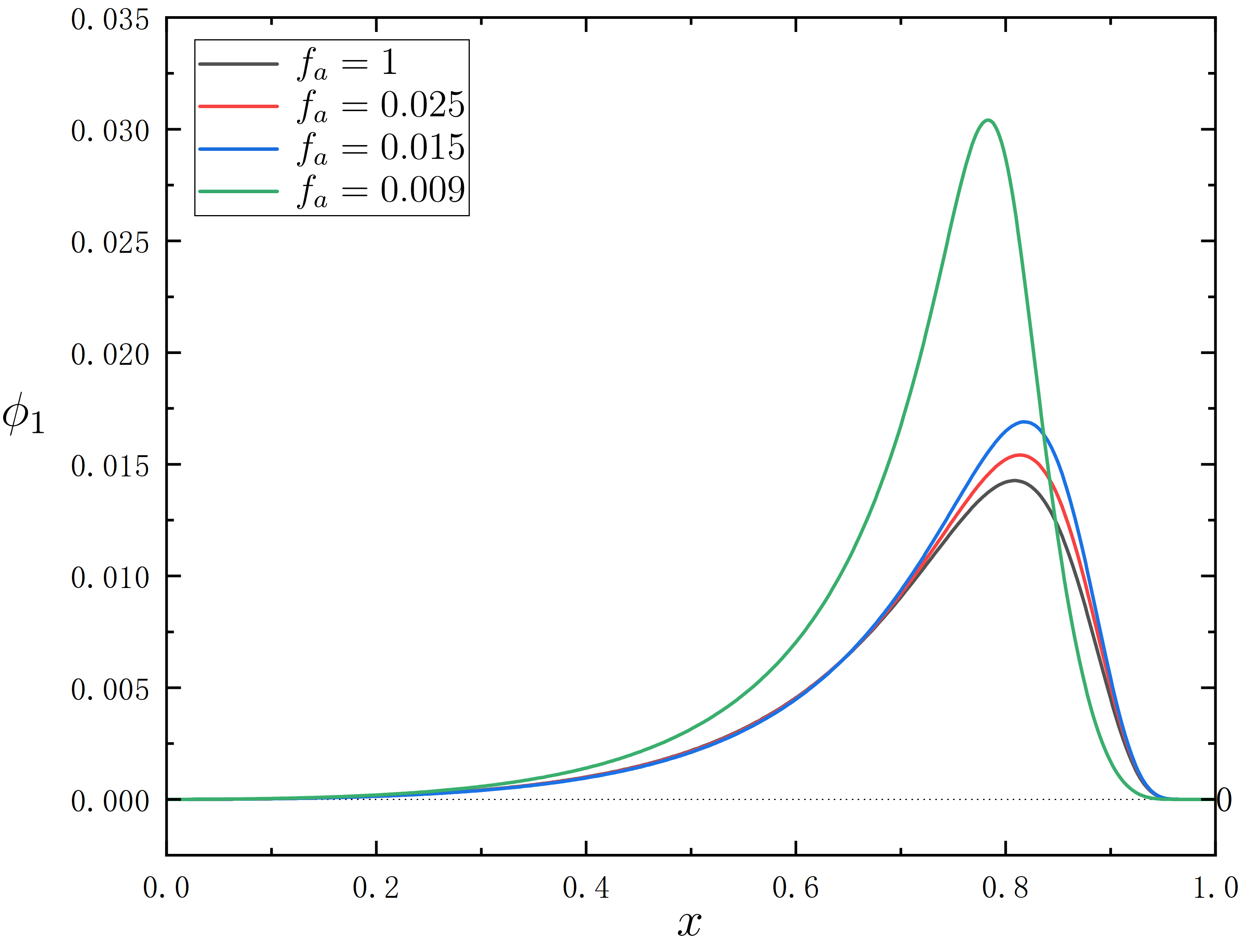} } 
\end{center}
\caption{The radial distribution of two axion fields in the RMABSs with the $^1S^2P$ state at $\theta=\pi/4$. \textit{Left}: The distribution of the axion field $\phi_0$ in the ground state $^1S$ state as a function of $x$ and $\theta$ for the frequency $\omega_0=0.8$. 
\textit{Right}: The distribution of the other axion field $\phi_1$ in the excited state $^2P$ state as a function of $x$ and $\theta$ for the frequency $\omega_1=0.8$.}
\label{fig:1s2p-distribution}
\end{figure}

In the last subsection, we gave a family of excited state $^2S$ and multistate $^1S^2S$ solutions, 
which means the value of the axion field $\phi_1$ in the $^2S$ state change the sign along the radial $r$ direction. 
In this subsection, we will show the numerical results of $^2P$ state and $^1S^2P$ state.
For $^1S^2P$ state, we plot the radial distributions of $\phi_0$ and $\phi_1$ in Fig.~\ref{fig:1s2p-distribution}. 
Although the two fields have no nodes in the radial direction, 
we know from prior work that the angular node is in the equatorial plane~\cite{Wang:2018xhw,Li:2019mlk}. 
The radial distribution of $\phi_0$ and $\phi_1$ becomes more compact. 

\begin{figure}[h!]
  \centering
      \includegraphics[width=0.9\textwidth]{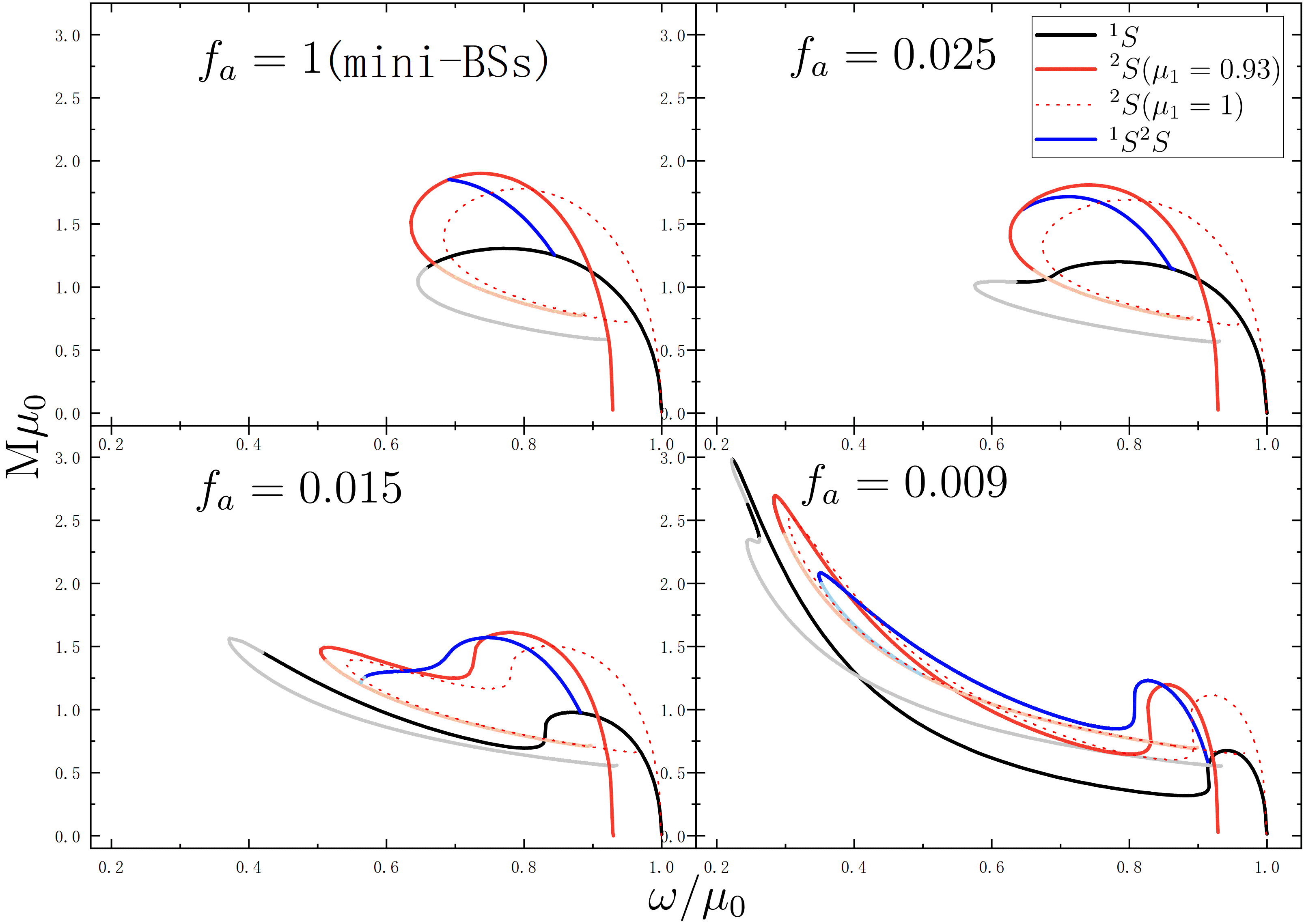}
      \caption{The ADM mass of the rotating axion stars as a function of the axion field frequency. The black line indicates $^1S$ state, the red solid line indicates the $^2P$ state with $\mu_1=0.93$, the red dotted line indicates the $^2P$ state with $\mu_1=1$, and the blue line indicates $^1S^2P$ state in the case of synchronized frequency, respectively. The light-colored line segment where the solutions exhibit an ergo-surface. }
      \label{M1s2p-sycronize-M}
\end{figure}

In Fig.~\ref{M1s2p-sycronize-M}, we plot ADM mass $M$ as a function of the frequency $\omega$ with the ground state $^1S$~(black line), the first excited state $^2P$ with $\mu_1=0.93$~(solid red line), the first excited state $^2P$ with $\mu_1=1$~(dotted red line), the co-existing state $^1S^2P$~(blue line), respectively. 
Here, the black line in Fig.~\ref{M1s2p-sycronize-M} and the black line in Fig.~\ref{M1s2s-sycronize-M} are identical, they represent the ground state $^1S$. 
The dotted red line starts from the vacuum~($\omega_1=1$), and spirals to the center. 
Similarly, the excited configuration with small sufficient $f_a$ has a higher maximum mass than the fundamental configuration. 
The $^2P$ state with $\mu_1=0.93$~(solid red line) has a higher ADM mass and a lower minimum frequency than The $^2P$ state with $\mu_1=1$. 
We also analyze the properties of the co-existing state $^1S^2P$ in the synchronized frequency condition. 
The co-existing state $^1S^2P$ connects the existence domain of $^1S$ state and $^2P$ state.  
As frequency increases or decreases, we can obtain $^1S$ state and $^2P$ state, respectively. 
The monotonic curve with $f_a=1$ is the simplest. 
For the solutions with sufficiently small $f_a$, the mass of $^1S^2P$ state may be higher than that of $^1S$ state and $^2P$ state. 
For the $^1S$ state, $^2S$ state, and $^1S^2P$ state, we find that the decrease of $f_a$ makes the minimum value of frequency lower. 
We also obtain the solutions with the second branch and ergo-surface. 
Starting from $\omega=1$, RABSs in the $^2P$ state firstly show ergo-surface near the minimum frequency. 

\begin{figure}[h!]
  \centering
      \includegraphics[width=0.9\textwidth]{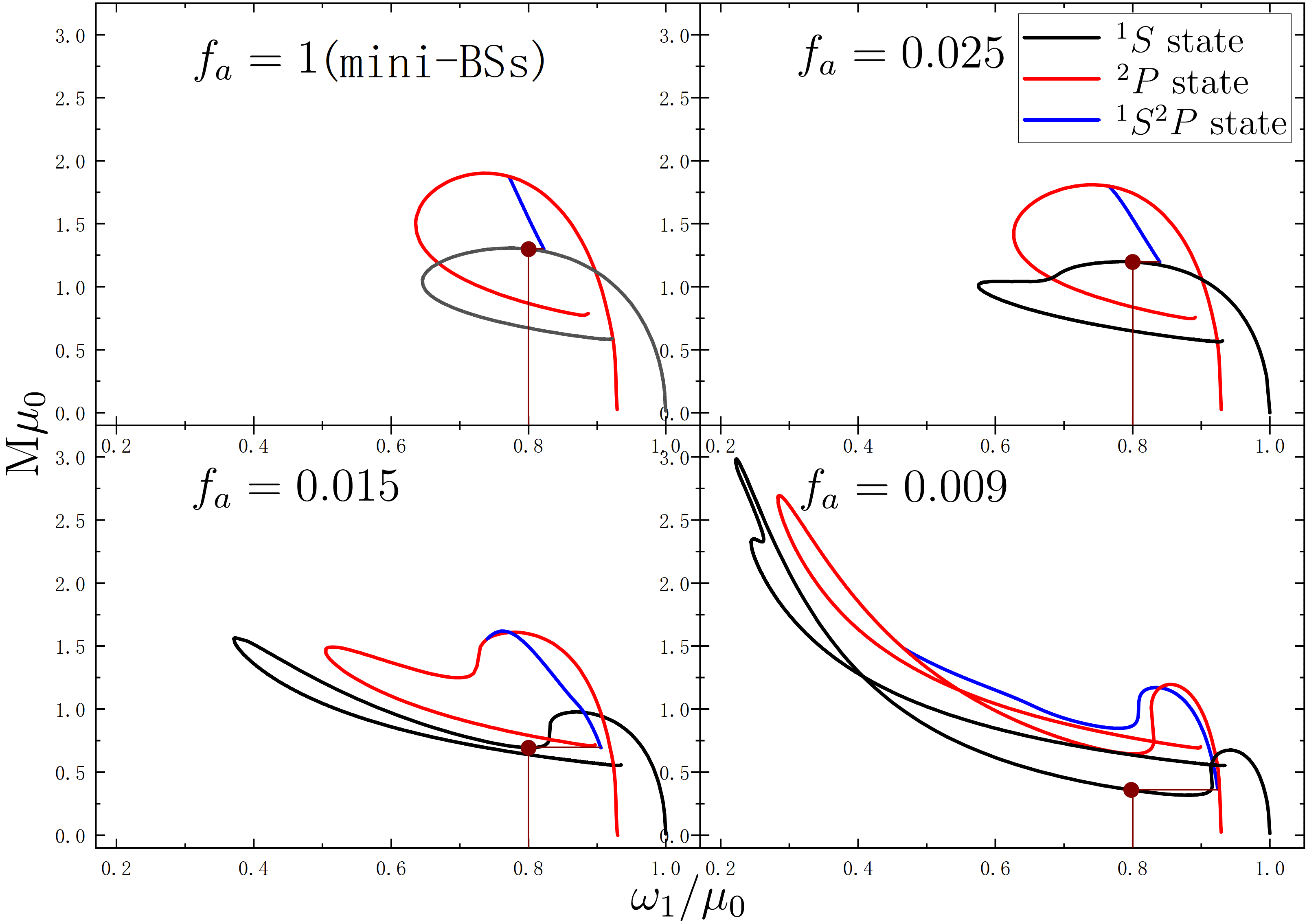}
      \caption{The ADM mass of the rotating axion stars as a function of the axion field frequency. The black line indicates $^1S$ state, the red solid line indicates the $^2P$ state with $\mu_1=0.93$, the red dotted line indicates the $^2P$ state with $\mu_1=1$, and the blue line indicates $^1S^2P$ state in the case of nonsynchronized frequency, respectively.}
      \label{M1s2p-nonsycronize-M}
\end{figure}

The co-existing state $^1S^2P$ in the nonsynchronized frequency condition is studied in Fig.~\ref{M1s2p-nonsycronize-M}. 
The difference between Fig.~\ref{M1s2p-sycronize-M} and Fig.~\ref{M1s2p-nonsycronize-M} is only the blue line. 
The frequency $\omega_0$ is not equal to the frequency $\omega_1$ under the nonsynchronized frequency condition. 
Here, we set $\omega_0=0.8$. 
When $\omega_1$ tends to the maximum or minimum value, the co-existing state $^1S^2P$ will reduce to the single axion field in the ground state $^1S$ with $\omega_0=0.8$ and the other axion field in the excited state $^2P$. 
Like the synchronized frequency condition, the rotating multistate axion stars in the nonsynchronized frequency condition also have no second branch. 

\begin{figure}[h!]
\begin{center}
\subfigure{\includegraphics[height=0.25\textheight]{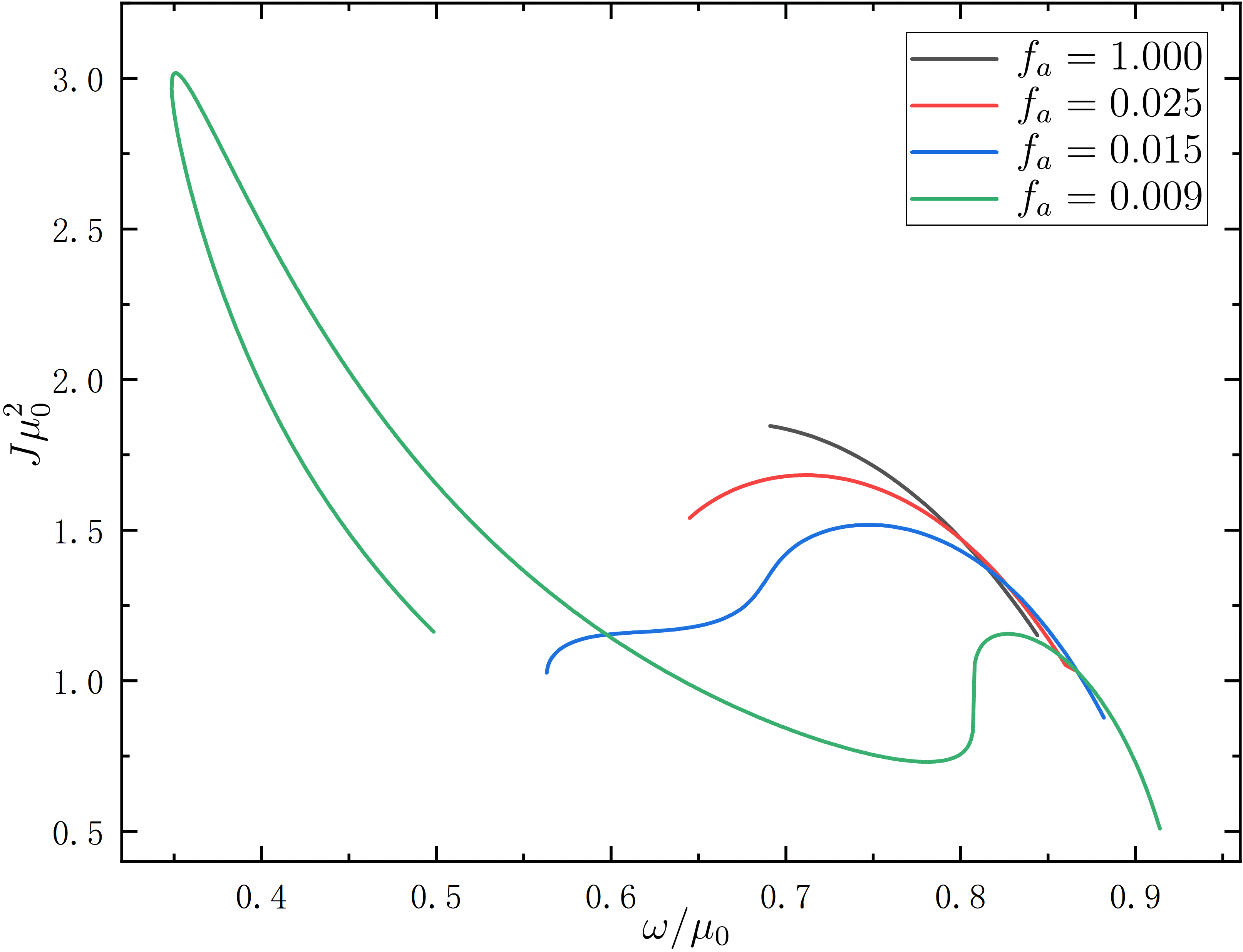} }
\subfigure{\includegraphics[height=0.25\textheight]{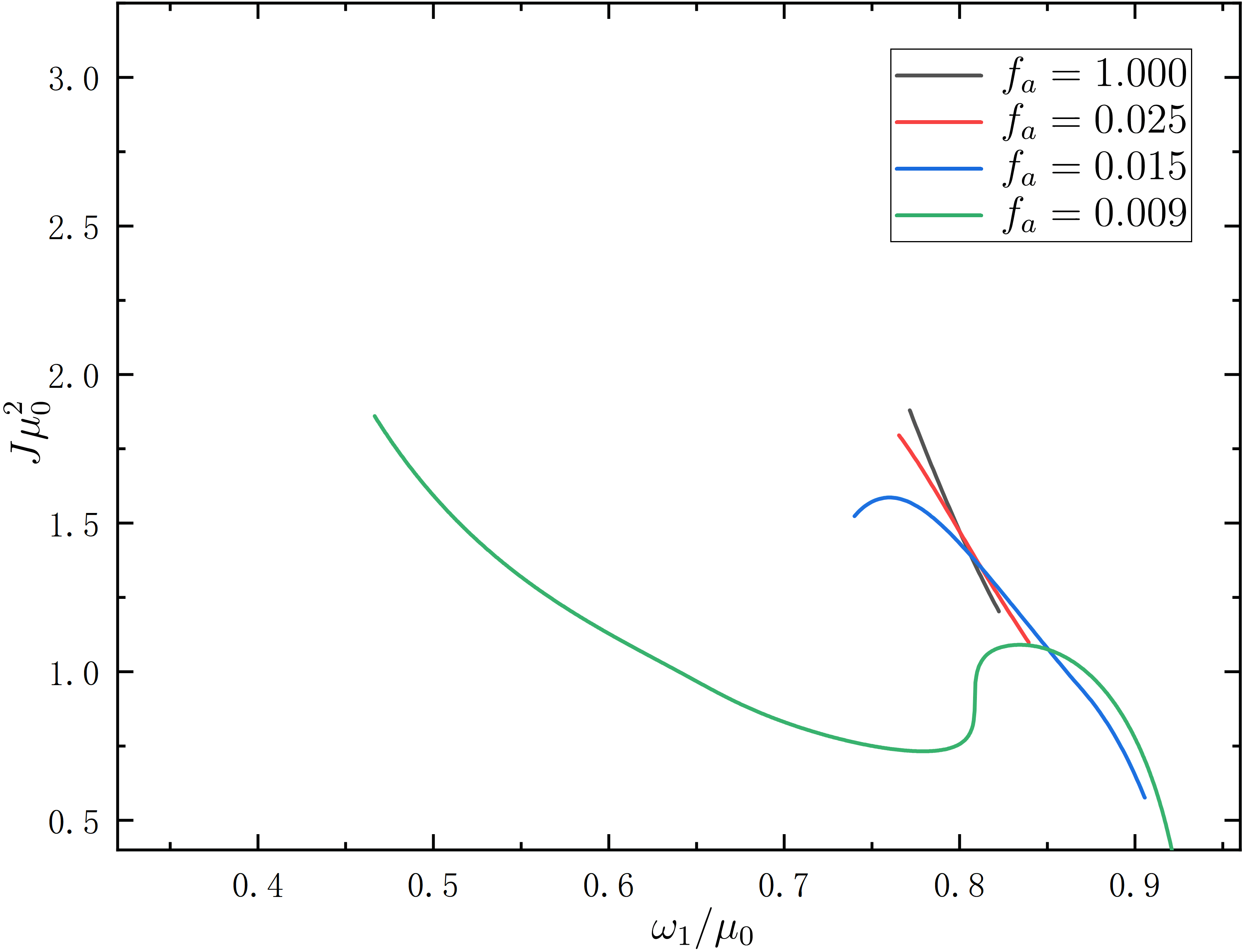} } 
\end{center}
\caption{The angular momentum of the RMABSs versus the axion field frequency for the coexisting state $^1S^2P$ with $f_a=\{1,0.025,0.015,0.009\}$. \textit{Left}: The RMABSs under the conditions of synchronized frequency. \textit{Right}: The RMABSs under the conditions of nonsynchronized frequency~($\omega_0=0.8$).}
\label{fig:1s2p-J}
\end{figure}

We also show $J-\omega$ diagram under the conditions of synchronized frequency and nonsynchronized frequency. 
On the left panel of Fig.~\ref{fig:1s2p-J}, the angular momentum of the case of $f_a=1$ versus synchronized frequency $\omega$ is monotonic. 
As $f_a$ decreases, the shape of curves becomes complicated. 
Unlike the $^1S^2S$ state, 
for $f_a=0.009$, the maximum angular momentum is much higher than the other three samples. 
The curves corresponding to $f_a=0.009$ present the second branch. 
On the right panel of Fig.~\ref{fig:1s2p-J}, all three curves are monotonic except for $f_a=0.009$. 
Moreover, the curves of $f_a=0.009$ still do not show the second branch, because the minimum frequency is not small enough. 

\begin{figure}[h!]
\begin{center}
\subfigure{\includegraphics[height=0.3\textheight]{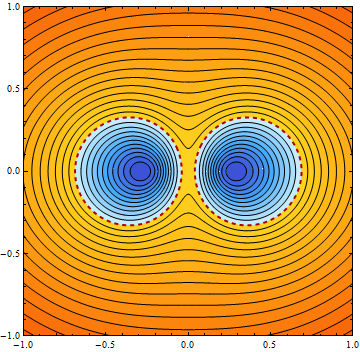} }
\subfigure{\includegraphics[height=0.3\textheight]{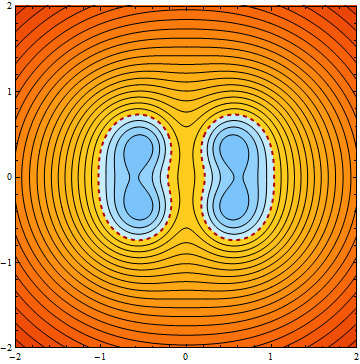} } 
\end{center}
\caption{The ergo-region structure of multistate configuration with $f_a=0.009$ in $(r,\theta)$ polar coordinates. The orange region means that $g_{tt}$ is positive. The blue region means that $g_{tt}$ is negative. The left panel indicates the $^1S^2S$ state with $\omega=0.7$~(second branch). The right panel indicates the $^1S^2P$ state with $\omega=0.49$~(second branch). }
\label{ergo-surface}
\end{figure}

Ergo-region is a key property for Kerr black hole. 
Due to $g_{tt} > 0$, a free particle in an ergo-region must rotate with the black hole if its geodesic remains time-like. 
Thus, this allows energy extraction from Kerr black hole~\cite{Penrose:1969pc,Penrose:1971uk}. 
Rotating mini-boson stars may also have an ergo-region~\cite{Cardoso:2007az}. 
The rotating axion stars solutions with an ergo-region is shown in Fig.~\ref{M1s2s-sycronize-M} and \ref{M1s2p-sycronize-M}. 
The ergo-region of rotating axion boson stars in the $^1S$ state has been studied in Ref.~\cite{Delgado:2020udb}. 
When $f_a$ is large enough, as the $M-\omega$ curve spirals into the center, ergo-region firstly appears near the minimum frequency. 
Then the solutions in the second branch always have the ergo-region. 
For fundamental solutions with smaller $f_a$, the solutions with an ergo-region may have two disconnected regions in the parameter space. 
For the excited configuration, all solutions with an ergo-region are connected. 
For the RMABSs under the synchronized frequency condition, the solutions with small $f_a$ allow the existence of ergo-region. 
This is not the same as the multistate configuration of rotating mini-boson stars~($f_a=1$). 
However, the RMABSs under the nonsynchronized frequency condition have no observed ergo-region. 
We show the ergo-region structure of the $^1S^2S$ state and $^1S^2P$ state in Fig.~\ref{ergo-surface}. 
They are solutions on the second branch. 
For the $^1S^2S$ state, the ergo-region is a standard torus.  
The ergo-region of the $^1S^2P$ state is a flat ring. 
The difference in shape between the two co-existing states comes from the symmetry of the excited axion field about the equatorial plane.

\section{Conclusions}\label{sec5}
In this work, we have constructed the first-excited configuration and multistate configuration of rotating axion boson stars. 
The first-excited RABSs have two types of solutions, including the even parity $^2S$ state and the odd parity $^2P$ state. 
The RMABSs are composed of the superposition of fundamental configuration and first-excited configuration. 
Thus the RMABSs also include the $^1S^2S$ state and the $^1S^2P$ state. 
For fundamental, first-excited and multistate configuration, we found that the existence domain of the solutions expands as the $f_a$ decreases. 
This allows the existence of lower frequency solutions. 
For single-field configurations, the minimum frequency corresponds to near the global maximum of the mass when the axion decay constant is small enough, and the maximum mass increases with the decrease of $f_a$. 
RABSs with sufficient small $f_a$ exhibit some properties that are different from rotating mini-boson stars~($f_a=1$). 
First, the fundamental configuration may have a heavier maximum mass than the excited configuration. 
Then, the ADM mass of the RMABSs can be higher than both the fundamental configuration and the excited configuration at the same frequency in some regions of parameter space. 
Finally, with the emergence of the second branch, the ergo-surface also emerges in the model of the RMABSs. 

How to understand the stability of the RABSs and the RMABSs is an interesting question. 
Firstly, the solutions with ergo-surface are plagued by superradiant instabilities. 
How about the dynamically stability of the solutions of without ergo-surface? 
The rotating mini-bosons without self-interaction are dynamically unstable, including the ground state and the excited state~\cite{Sanchis-Gual:2019ljs,DiGiovanni:2020ror,Sanchis-Gual:2021edp}. 
However, the sufficiently strong self-interactions can stabilize boson stars. 
In Ref.~\cite{Siemonsen:2020hcg}, axionic potential with the sufficiently small $f_a$ can quench the instability of the fundamental solutions of rotating axion boson star. 
The authors of the Ref.~\cite{Sanchis-Gual:2021phr} also found that the sufficiently strong quartic self-interactions also stabilize spherical excited boson stars, which can be formed from dilute initial data. 
Here, we do not know if the strong axion potential can help stabilize the excited rotating axion boson stars. 
For RMABSs, some studies show that the stability of the multistate boson stars depends on the stability of the components~\cite{Bernal:2009zy,Sanchis-Gual:2021edp}. 
Adding a stable fundamental boson star to an unstable BS is a kind of stabilization mechanism against decaying to a fundamental configuration. 
We may expect that the RMABSs also be stable if at least one of two components is dynamically stable. 
Nevertheless, we will further investigate the stability issue of RMABSs in the future. 

\section*{Acknowledgements}
This work is supported by National Key Research and Development Program of China (Grants No.~2020YFC2201503 and No. 2022YFC2204100) and  the National Natural Science Foundation of China (Grants No.~12275110 and No.~12247101).

\providecommand{\href}[2]{#2}\begingroup\raggedright

\endgroup
\end{document}